\documentclass[sigconf]{acmart}

\renewcommand\footnotetextcopyrightpermission[1]{} 

\begin{document}

\title{IOPathTune: Adaptive Online Parameter Tuning for Parallel File System I/O Path}

\author{Md. Hasanur Rashid}
\email{mrashid2@uncc.edu}
\affiliation{%
  \institution{University of North Carolina at Charlotte}
  \country{Charlotte, NC, USA}
}
\author{Youbiao He}
\email{yh54@iastate.edu}
\affiliation{%
  \institution{Iowa State University}
  \country{Ames, IA, USA}
}
\author{Forrest Sheng Bao}
\email{forrest.bao@gmail.com}
\affiliation{%
  \institution{Iowa State University}
  \country{Ames, IA, USA}
}
\author{Dong Dai}
\email{ddai@uncc.edu}
\affiliation{%
  \institution{University of North Carolina at Charlotte}
  \country{Charlotte, NC, USA}
}

\renewcommand{\shortauthors}{Rashid et al.}

\begin{abstract}
    \textbf{Motivations.}
    Parallel file systems (PFSes) play a foundational role in High-Performance Computing (HPC) platforms for providing users and applications the data access ability at needed speed and capacity. It is critical to make sure PFSes could deliver the extreme performance for a largely diverse spectrum of scientific applications running in HPC systems. Parallel file systems are known for their long and complex I/O path. Specifically, in PFSes, I/O requests are issued from the computing nodes or I/O nodes using PFS client library. They then are delivered to the storage servers using RPC mechanisms over high-speed network. Finally, these requests will be buffered on the storage servers' I/O queues waiting for the internal I/O schedulers to materialize the data. An efficient and ideal I/O path means the data flows from clients to storage servers without being jammed at any places, which is then a key to deliver optimized I/O performance in PFSes. To achieve optimal I/O performance, the parallel file systems must have proper settings on multiple PFS parameters that control how data flows in the I/O path. Although carefully picked, the default settings often fail to deliver optimal performance for diverse and changing workloads. 
    
    To effectively tune I/O path-relevant parameters for parallel file systems, we look for several key properties: adaptive, timely, and flexible. The tuning framework should adjust parameters adaptively when workloads change and respond to runtime variables such as I/O contentions. The tuning decision delivery needs to be fast due to the short burst nature of the workloads. The tunable parameters should also be modifiable dynamically. The framework must accommodate the flexibility to tune different clients differently. Due to the limited I/O, computation, and communication resources for performance tuning, the framework should also avoid the expensive probing or profiling of the HPC environment.
    
    \noindent\textbf{Design of IOPathTune.} 
    In this study, we propose IOPathTune, a new tuning framework designed to tune Lustre \cite{braam2019lustre} I/O Path online from the client side adaptively. We focus on two parameters after closely investigating the Lustre I/O path: \texttt{max\_pages\_per\_rpc} (maximum number of pages in a single RPC) and \texttt{max\_rpcs\_in\_flight} (maximum number of RPCs in flight). Both of them belong to the Lustre Portal RPC (PtlRPC) subsystem. They are dynamically tunable, take immediate effect, and control the I/O flow for each Lustre Client. These important features align well with our goals. Due to these two parameters, the RPCs are formed and transferred in a structured way, helping us avoid the characterization of workloads. The stat data associated with these two parameters are also readily available and require no extra system probing. 
    
    \begin{figure}[htbp]
    \centering
    \includegraphics[width=0.45\textwidth]{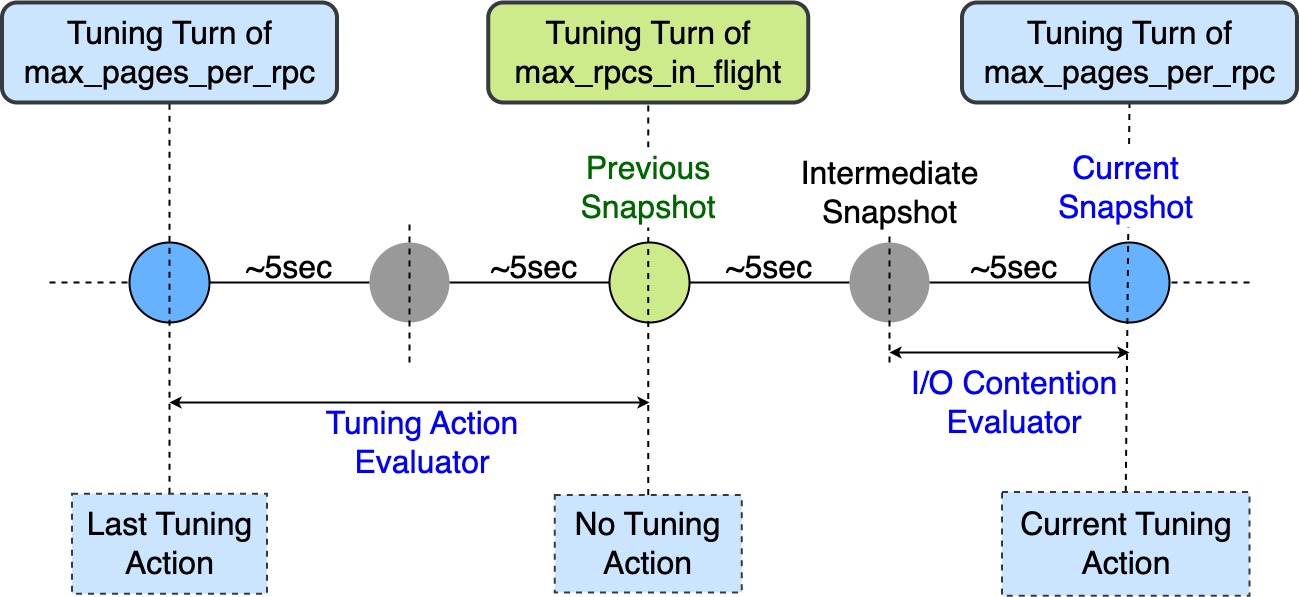}
    \caption{Heuristic Tuning Approach}
    \label{fig:multi_snap}
    \end{figure}
    
    One critical design choice of IOPathTune is to avoid probing storage servers or other compute nodes. We design it to solely depend on the statistics collected by the PFS client library. Our tuning algorithm proceeds with tuning every ten seconds, and in each turn, it chooses to tune either of the two parameters alternately as illustrated in Figure \ref{fig:multi_snap}. The tuning action consists of either multiplying or dividing the parameter value by two each time, much like the primary approach of TCP congestion control. 
    
    Specifically, during tuning, we calculate different metrics along the I/O path. The metrics we derive are: how much data is in the dirty page cache, how fast the pages are getting cached, how quickly the client can generate RPCs, and what speed the client is achieving while transferring RPCs for the last observation period. By comparing these metrics, if we observe that our last tuning action improves the bandwidth, we reciprocate the previous tuning action. Otherwise, we do the opposite of the last tuning action. We also observe whether I/O contentions are developing or not based on these metrics. In the case of I/O contentions, we adopt a conservative approach: assign blame on our previous tuning action and revert it.

    \begin{table*}[htpb]
    \caption{Standalone Workload Executions}
    \begin{center}
    \begin{tabular}{|c|c|c|c|}
    \hline
    {} & {\textbf{8KB I/O Request}} & {\textbf{1MB I/O Request}} & {\textbf{16MB I/O Request}}\\
    \cline{2-4} 
    \textbf{Workload} & \textbf{\textit{I/O Throughput}} & \textbf{\textit{I/O Throughput}} & \textbf{\textit{I/O Throughput}}\\
    \textbf{Name} & \textbf{\textit{Improvement(\%)}} & \textbf{\textit{Improvement(\%)}} & \textbf{\textit{Improvement(\%)}}\\
    \hline 
    Random Write & 7.82 & \textbf{22.97} & 10.93\\
    \hline 
    Fivestream Random Write & \textbf{64.46} & \textbf{231.98} & \textbf{43.44}\\
    \hline 
    Random Read-Write & -7.46 & 5.57 & -2.91\\
    \hline
    Sequential Write & -4.39 & -0.73 & 7.56\\
    \hline 
    Fivestream Sequential Write & -7.29 & 3.75 & -7.59\\
    \hline 
    Sequential Read-Write & 4.03 & \textbf{113.19} & \textbf{72.6}\\
    \hline
    Whole File Write & {} & {} & \textbf{86.45}\\
    \hline 
    Whole File Read-Write & {} & {} & \textbf{96.58}\\
    \hline
    \end{tabular}
    \label{tab1}
    \end{center}
    \end{table*}

    \begin{table*}[htbp]
    \caption{Multiple Client Executions}
    \begin{center}
    \begin{tabular}{|c|c|c|c|c|}
    \hline
    \textbf{Workload} & \textbf{Client} & {\textbf{Default Execution}} & {\textbf{CAPES Execution}} & {\textbf{Heuristic Execution}}\\
    \cline{3-5} 
    \textbf{Name} & \textbf{Name} & \textbf{\textit{I/O BW(MB/s)}} & \textbf{\textit{I/O BW(MB/s)}} & \textbf{\textit{I/O BW(MB/s)}}\\
    \hline 
    Fivestream Random Write & node1 & 385.4 & 237 & \textbf{2627.9}\\
    \hline 
    Random Write & node2 & 95.2 & 101.4 & \textbf{206.3}\\
    \hline 
    Random Read-Write & node3 & 2127.6 & \textbf{4209.3} & 3199.8\\
    \hline 
    Sequential Read-Write & node4 & 639.2 & 630.8 & \textbf{1134.6}\\
    \hline 
    Whole File Read-Write & node5 & 1682.3 & 784.3 & \textbf{4135}\\
    \hline
    \multicolumn{2}{|c|}{\textbf{Total Multi-client BW (MB/s)}} & {4929.7} & {5962.8} & {\textbf{11303.6}}\\
    \hline
    \end{tabular}
    \label{tab2}
    \end{center}
    \end{table*}
    
    \noindent\textbf{Evaluations.} 
    To test the efficacy of our framework IOPathTune, we leverage CloudLab \cite{duplyakin2019design} to conduct the evaluations. Our HPC cluster consists of one machine for MGS/MDS, four for OSS, and five for clients (with an extra machine while running CAPES). Every machine has two Intel Xeon Silver 4114 10-core CPUs at 2.20 GHz, 192GB ECC DDR4-2666 memory, and one Intel DC S3500 480 GB 6G SATA SSD. Each of our OSS consists of two 200GB OSTs mounted on SSD. Dual-port Intel X520-DA2 10Gb NIC supports the network. We have set up CentOS 7 as the operating system and Lustre 2.12.8 as the primary file system on our HPC cluster. We tested 20 different Filebench \cite{tarasov2016filebench} workloads with varying I/O patterns (random or sequential), I/O operations (read or write), I/O request sizes, number of processes and threads, and others. We ran three different evaluations and checked how IOPathTune handles different scenarios.
    
    The first test was standalone workload executions from a single client. We observed either on par with slight degradation or better performance than the default configuration across all workloads. Table \ref{tab1} shows some of the most considerable improvement includes 231\%, 113\%, and 96\% for Filebench fivestreamwriternd, whole file read-write, and sequential read-write workloads. Our second kind of test was dynamic testing, where we continuously changed the workloads after running for 300 seconds. We changed the workload six times at a single run and experimented with five runs, each with six different combinations of workloads. We observed consistent improvements from IOPathTune similar to the standalone performance evaluation, indicating our algorithm can quickly catch up and converge to better configurations.
    
    Our final type of test was executing different workloads from different clients at the same time. We compared IOPathTunes' performance with the default configuration and CAPES \cite{li2017capes} execution. IOPathTune improved the total bandwidth of the cluster by 129.30\% compared to the default execution and by 89.57\% compared to the CAPES execution as shown in Table \ref{tab2}. The findings prove that our proposed IOPathTune framework can independently tune parameters from different clients to achieve better performance.
    
    \noindent\textbf{Conclusions.}
    Our study has shown how we can perform adaptive online parameter tuning without characterizing the workloads, without doing expensive profiling, and without communicating with other machines. We hope this approach will welcome more attention towards researching inexpensive yet very effective solutions in system research. In future, we like to scale our algorithm to accommodate tuning more parameters following this heuristic approach. We would also like to test it out in real-world HPC facilities to observe how much improvement the solution brings regarding I/O performance.
    
    \noindent\textbf{Acknowledgement.}
    This research is partially supported by NSF award CNS-1817089 and CNS-2008265.
\end{abstract}


\maketitle
\pagestyle{plain} 

\bibliographystyle{ACM-Reference-Format}
\bibliography{software}

\end{document}